\documentclass[aps,prl,reprint,superscriptaddress,longbibliography,floatfix]{revtex4-2}
\usepackage[T1]{fontenc}
\usepackage[utf8]{inputenc}
\usepackage{lmodern}
\usepackage{amsmath,amssymb,graphicx,bm,microtype}
\usepackage[hidelinks]{hyperref}
\newcommand{\girr}{\gamma_{\mathrm{irr}}}
\newcommand{\gvis}{\gamma_{\mathrm{vis}}}
\newcommand{\one}{\mathbf 1}
\newcommand{\E}{\mathbb E}
\newcommand{\KL}{D_{\mathrm{KL}}}
\newcommand{\Rnull}{\mathcal R_{\mathrm{hidden}}}
\begin{document}
\title{The Spectral Cost of Detecting Nonequilibrium at Finite Temporal Resolution}
\author{Murad Aznagulov}
\email{amurad12@icloud.com}
\affiliation{Faculty of Physics, Lomonosov Moscow State University, Moscow, Russia}
\date{September 2026}
\begin{abstract}
A detector that cannot take two snapshots closer than a time $\tau_0$ still identifies a fully observed finite Markov generator exactly, provided its gaps are randomized by two clocks. Identification is not detection. For a normal nonequilibrium network, the number of snapshots needed to reject every hidden equilibrium explanation grows as $e^{2\gamma_{\mathrm{irr}}\tau_0}$, where $\gamma_{\mathrm{irr}}$ is the slowest damping of rotating modes. No adaptive schedule beats this rate, and a test on one dependent trajectory attains it.
\end{abstract}
\maketitle

% CORE-BEGIN
Stochastic systems are rarely watched continuously. A camera has a frame time, a photon counter a dead time, an amplifier a rise time. A molecular machine that completes its dissipative cycle faster than the instrument can look twice may appear balanced from one snapshot to the next. Two questions then come apart. Do the data still determine the underlying dynamics? And how many data are needed to tell that the system is out of equilibrium? We show that the answers diverge. The dynamics remains exactly identifiable at any finite resolution, yet detecting nonequilibrium costs exponentially many snapshots in the minimum gap, at a rate fixed by the irreversible part of the spectrum.

An exponential cost is what a signal-to-noise estimate suggests: irreversible signals decay with the gap, fluctuations do not. Such an estimate, however, concerns one estimator and one schedule. The result here is stronger on both sides. The lower bound holds for every causal timing strategy and stopping rule, even one that knows the competing models exactly. The upper bound is attained by a fixed protocol on a single uninterrupted trajectory, and its false-positive rate is controlled against \emph{every} finite hidden equilibrium model, with no bound on hidden dimension or mixing time. The rate is set by the slowest damping of rotating modes, which can be unrelated to the longest relaxation time.

Thermodynamic inference extracts dissipation from incomplete records \cite{Seifert2012,SkinnerDunkel2021,vanDerMeerErtelSeifert2022}, and the role of coarse graining and imperfect resolution has been studied for estimators and dissipative time scales \cite{CisnerosEtAl2023,YuHarunari2024,FritzEtAl2025}. Stroboscopic reconstruction \cite{BauerEtAl} and the compensation of detection blackouts \cite{MaierEtAl2026} treat related limitations of the record. Exponential decay of correlation asymmetry \cite{VuVoSaito2024}, spectral windows for testing after a known reversible channel \cite{BerthetKanade2019}, and sequential or goodness-of-fit tests of time direction and Markov dynamics \cite{RoldanEtAl2015,SteuberEtAl2012,BesagMondal2013,WolferKontorovich2020} are close precedents. What is new is the optimal cost of the measurement itself under a hard minimum gap, together with a protocol that attains it against an unknown hidden equilibrium mechanism.

\emph{Observation model.---}Let $Q$ be a finite irreducible generator in row convention, with stationary law $\pi>0$ and ordinary time reversal. Write
\begin{equation}
 Q^\dagger=D_\pi^{-1}Q^\top D_\pi,\quad
 S=\tfrac12(Q+Q^\dagger),\quad A=\tfrac12(Q-Q^\dagger).
 \label{eq:parts}
\end{equation}
Detailed balance means $Q=Q^\dagger$. The reversible generator $S$ has the same $\pi$, the same exit rate from every state and the same stationary activity as $Q$.

Only the states at measurement times are recorded. Each gap $T_k\ge\tau_0>0$ is chosen from the past record and private randomness before the next state is seen; the times may be stored, and stopping may depend on the record. There is no jump counter, integrated signal or intervention between snapshots, and the process starts in stationarity.

The null class $\Rnull$ contains the stationary observations, on the same alphabet, of all finite reversible Markov chains with deterministic labels; their dimension, rates and mixing times are arbitrary. For a fully observed irreversible alternative $Q$, let $N_*(Q,\tau_0,\alpha)$ be the least expected number $\E_QN$ of observed transitions over tests whose false-positive probability is at most $\alpha$ for every member of $\Rnull$ and whose false-negative probability under $Q$ is at most $\alpha$. This is an instance-wise optimum for a fixed $Q$, not a worst case over generators.

\begin{figure}[tbp]
\includegraphics[width=\columnwidth]{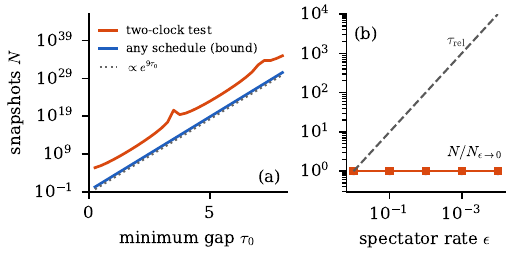}
\caption{Observation cost for the driven ring $G$ of Eq.~\eqref{eq:product} at $\alpha=0.01$. (a)~Blue: rigorous analytic information lower bound valid for every causal schedule and stopping rule; for the plotted range $\max_i\KL[(e^{tG})_{i\cdot}\Vert(e^{tS})_{i\cdot}]\le24e^{-9t}$. Orange: deterministic 60-digit central-limit power budget of the two-clock current test ($r_\ell=1,4$), shown as a protocol diagnostic rather than the theorem's conservative finite-budget constant. Both have exponent $9=2\girr$. (b)~For the product chain $Q_\epsilon$ at $\tau_0=1$, the lower bound uses only the driven factor and the test ignores the reversible spectator, exactly as in the proof; both observation costs are therefore independent of $\epsilon$ while the relaxation time $1/(2\epsilon)$ grows by four orders of magnitude.}
\label{fig:times}
\end{figure}

\emph{Sharp spectral law.---}Call $Q$ normal if $QQ^\dagger=Q^\dagger Q$, equivalently $[S,A]=0$: relaxation and circulation commute, and the dynamics splits into purely relaxing modes and independently damped rotating ones. Let
\begin{equation}
 \girr=\min_{\substack{\lambda\in\operatorname{spec}Q\\ \operatorname{Im}\lambda\ne0}}
              (-\operatorname{Re}\lambda).
 \label{eq:rate}
\end{equation}
For every irreversible normal $Q$ there are constants $c_Q,C_Q>0$ and $\tau_Q<\infty$ such that, for $\tau_0\ge\tau_Q$ and $0<\alpha\le1/4$,
\begin{equation}
 c_Q e^{2\girr\tau_0}\log(1/\alpha)
 \le N_* \le
 C_Q e^{2\girr\tau_0}\log(1/\alpha).
 \label{eq:main}
\end{equation}
Hence $\tau_0^{-1}\log N_*\to2\girr$ (Fig.~\ref{fig:times}a): every additional irreversible damping time $1/\girr$ inside the gap multiplies the required record length by $e^2\approx7.4$. The factor $2$ is the usual quadratic relation between small probability differences and distinguishability, applied to mode amplitudes that decay as $e^{-\girr\tau_0}$. Normality is used only for the \emph{matching} exponent; the adaptive lower bound and the hidden-equilibrium guarantee hold without it. Translation-invariant walks on finite Abelian groups, and their products with reversible chains, are normal; the general case is treated below.

\emph{No adaptive shortcut.---}Compare $Q$ with the admissible equilibrium model $S$. With $\lambda_S$ the spectral gap of $-S$ and $\Pi=\one\pi$, contraction in $L^2(\pi)$ gives $\|e^{tQ}-\Pi\|_\pi\le e^{-\lambda_St}$ and $\|e^{tQ}-e^{tS}\|_\pi\le2e^{-\lambda_St}$. Once the transition probabilities of $S$ exceed fixed fractions of their stationary values, the quadratic bound on relative entropy yields
\begin{equation}
 \KL\bigl((e^{tQ})_{i\cdot}\Vert(e^{tS})_{i\cdot}\bigr)
 \lesssim_Q e^{-2\gamma t}
 \label{eq:rowkl}
\end{equation}
with $\gamma=\lambda_S$. For normal $Q$ the two semigroups coincide on the relaxing modes, and on a rotating block they differ by $e^{-\gamma' t}(e^{\omega tJ}-I)$ with $J^\top=-J$, $J^2=-I$; then Eq.~\eqref{eq:rowkl} holds with $\gamma=\girr$.

Under identical past records the timing policy acts identically under $Q$ and $S$, so the choice of the next gap carries no information of its own. The chain rule for relative entropy with a causal stopping time \cite{Kaufmann2016} bounds the information in the whole record by a constant times $e^{-2\gamma\tau_0}\E_QN$, while a decision with both errors below $\alpha$ needs information of order $\log(1/\alpha)$. Therefore
\begin{equation}
 \E_QN\gtrsim_Q e^{2\gamma\tau_0}\log(1/\alpha).
 \label{eq:lower}
\end{equation}
Because $S\in\Rnull$, this limits every procedure, not a particular estimator; knowing all exit rates and the total activity does not help, since $S$ shares them.

\emph{Two clocks.---}Fix distinct rates $r_1,r_2>0$. Before each snapshot draw a fair mark $\ell$, wait $\tau_0+\operatorname{Exp}(r_\ell)$, and record the mark with the state. Every gap respects the detector limit, and the sampled kernels are
\begin{equation}
 B_\ell=e^{\tau_0Q}r_\ell(r_\ell I-Q)^{-1}.
 \label{eq:clocks}
\end{equation}
Random-time sampling has a long history in estimation \cite{DuffieGlynn2004}. Here the common delayed evolution drops out: with $V=(r_2/r_1)B_1B_2^{-1}=(r_1-Q)^{-1}(r_2-Q)$,
\begin{equation}
 Q=(V-I)^{-1}(r_1V-r_2I),
 \label{eq:identify}
\end{equation}
for any finite generator, since $V-I=(r_2-r_1)(r_1-Q)^{-1}$ is invertible. Moreover, $Q$ is reversible exactly when both $B_\ell$ are: the kernels commute, and Eq.~\eqref{eq:identify} is then self-adjoint. Two clocks therefore keep the generator identifiable at every finite delay. The inversion is exponentially ill-conditioned, and the test below never performs it.

\emph{A guarantee against hidden equilibrium.---}Draw the number of transitions $K\sim\operatorname{Poisson}(M)$ in advance and independently; this randomizes how many gaps are used, not their minimum length. For a prespecified antisymmetric score $g(\ell,y,z)=-g(\ell,z,y)$ with $|g|\le1$, let $C_g$ be its sum over the $K$ recorded transitions. For every hidden equilibrium model,
\begin{equation}
 \E_{\rm eq}e^{\theta C_g}\le\exp[M(\cosh\theta-1)].
 \label{eq:mgf}
\end{equation}
No hidden dimension, rate or mixing time enters, although the visible process need not be Markovian.

The proof lifts $g$ to pairs of hidden states and forms $T_\theta(i,j)=\frac12\sum_\ell B_\ell(i,j)e^{\theta g(\ell,h(i),h(j))}$. Detailed balance gives $T_\theta^\dagger=T_{-\theta}$, so the symmetric part of $T_\theta$ is entrywise dominated by $\cosh\theta$ times a stochastic self-adjoint kernel and has largest eigenvalue at most $\cosh\theta$. Poissonization turns the moment-generating function into $\langle\one,e^{M(T_\theta-I)}\one\rangle_\pi$, and the logarithmic-norm bound for this semigroup gives Eq.~\eqref{eq:mgf} with all correlations between successive snapshots retained. Tilted-operator bounds of this kind are established for trajectory observables \cite{BakewellSmithEtAl2023}.

For each clock and each unordered pair of observed states, the test counts forward minus backward transitions. Chernoff's inequality with a union bound gives a common threshold of order $\sqrt{M\log(d^2/\alpha)}$ for $d$ observed states; the explicit finite-budget form is in the Supplemental Material \cite{Supplement}. The test uses neither the hidden mechanism nor its mixing time. Its guarantee covers the prespecified clocks, scores and Poissonized count, and not post hoc choices.

\emph{Attaining the rate.---}On a rotating block $-\gamma I+\omega J$,
\begin{equation}
 (B_\ell-B_\ell^\dagger)_{\rm block}
 =2e^{-\gamma\tau_0}
   [a_\ell\sin(\omega\tau_0)+b_\ell\cos(\omega\tau_0)]J,
 \label{eq:signal}
\end{equation}
with $(a_\ell,b_\ell)=r_\ell(r_\ell+\gamma,\omega)/[(r_\ell+\gamma)^2+\omega^2]$. Since $a_1b_2-a_2b_1\propto\omega(r_1-r_2)\ne0$, the two clocks respond with different phases and cannot both miss the signal at an unlucky delay. Orthogonality of the normal modes then gives a marked current with stationary mean
\begin{equation}
 |\mu|\gtrsim_Q e^{-\girr\tau_0}.
 \label{eq:mean}
\end{equation}

\begin{figure}[tbp]
\includegraphics[width=\columnwidth]{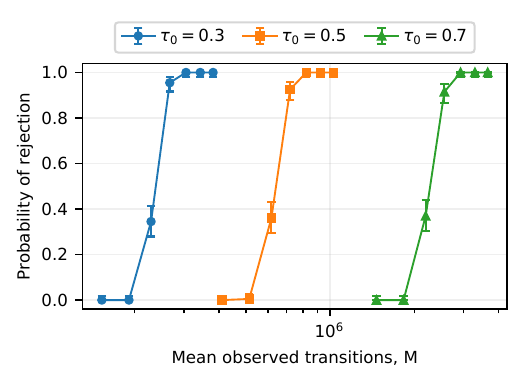}
\caption{Power of the one-record current test for the driven ring at $\alpha=0.01$. Bars are exact 95\% binomial intervals over independent whole records \cite{ClopperPearson1934}; snapshots within a record are dependent. The prespecified study shows the shift to larger budgets as $\tau_0$ grows; it is not a fit of the asymptotic exponent.}
\label{fig:power}
\end{figure}

For long delays the sampled chain contracts uniformly. A Poisson-equation decomposition writes each centered current as a martingale plus a bounded boundary term, and martingale concentration together with the tail of $K$ shows that $M$ of order $\mu^{-2}\log(1/\alpha)$ suffices to cross the threshold with probability $1-\alpha$. With Eq.~\eqref{eq:mean} this is the upper half of Eq.~\eqref{eq:main}, on one dependent trajectory. The test checks both signs of every current and needs no knowledge of which mode carries the signal. Explicit constants are given in the Supplemental Material \cite{Supplement}; for example, a budget $M=\lceil8192\,\eta_Q^{-2}e^{2\girr\tau_0}\log(4m/\alpha)\rceil$, with $\eta_Q$ a computable spectral constant and $m$ the number of scores, controls both errors.

\emph{Irreversible, not global, time scale.---}Take the three-state ring $G$ with clockwise rate $2$ and counterclockwise rate $1$, and attach an independent reversible coordinate,
\begin{equation}
 Q_\epsilon=G\otimes I_2+I_3\otimes
 \begin{pmatrix}-\epsilon&\epsilon\\\epsilon&-\epsilon\end{pmatrix}.
 \label{eq:product}
\end{equation}
For $0<\epsilon<9/4$ this normal chain produces entropy at rate $\log2$, has global gap $2\epsilon$, and has $\girr=9/2$ (Fig.~\ref{fig:times}). Its optimal count is $\Theta(e^{9\tau_0}\log(1/\alpha))$ with constants uniform in $\epsilon$ (Fig.~\ref{fig:times}b). The slow coordinate adds no information in the lower bound and is ignored by the test, so an arbitrarily slow equilibrium degree of freedom cannot make the fast cycle easier to detect. Equation~\eqref{eq:main} thus gives an operational meaning to the distinction between relaxation and dissipative scales \cite{CisnerosEtAl2023,YuHarunari2024}.

\emph{Beyond normal generators.---}For general $Q$ the same arguments give a window. Let $P_\lambda$ be the Riesz projector of an eigenvalue $\lambda\ne0$ and $N_\lambda=(Q-\lambda)P_\lambda$. Call $\lambda$ irreversibly visible if some generalized spectral coefficient carries antisymmetric stationary flux, $\operatorname{As}(D_\pi N_\lambda^kP_\lambda)\ne0$ for some $k$ below the Jordan index, and let $\gvis$ be the slowest damping among such eigenvalues. Then
\begin{equation}
 e^{2\lambda_S\tau_0}\lesssim_Q N_*/\log(1/\alpha)\lesssim_Q \tau_0^{-2k_*}e^{2\gvis\tau_0}
 \label{eq:window}
\end{equation}
when the slowest visible level is a single real eigenvalue or a single complex pair; here $k_*$ is the largest visible Jordan order, and $k_*=0$ in the diagonalizable case \cite{Supplement}. Always $\lambda_S\le\gvis$, because the numerical range of $Q$ on mean-zero functions lies in $\operatorname{Re}z\le-\lambda_S$. For normal $Q$, real eigenvalues carry no flux, $k_*=0$, and $\gvis=\girr$, which closes the exponential window. In nonnormal networks, by contrast, even a purely real spectrum can carry a detectable irreversible signal.

\emph{Finite data.---}In prespecified simulations every recorded state is the initial state of the next gap; there are no resets and no independent-pair approximation within a record. Figure~\ref{fig:power} shows the power curves moving to larger budgets as $\tau_0$ increases, while hidden equilibrium controls, including a slowly mixing one, stay within the nominal false-positive level. All budgets, repetitions, intervals and rescaled curves are reported in the Supplemental Material. The simulations assess a conservative implementation rather than the optimal constant; delay-dependent phases remain visible after exponential rescaling.

The results assume stationary initialization, passive state snapshots, ideal timing and the stated randomized count. A fully observed alternative is essential for the rate: sufficiently coarse observations of a driven system can carry no current at all. Recording time scales with the same exponential and an extra factor asymptotic to $\tau_0$.

Randomized timing thus preserves exact identifiability at any finite resolution without making nonequilibrium cheap to detect. Within passive snapshot measurements with a hard minimum gap, no causal schedule avoids the exponential loss of irreversible information, and for normal networks its rate is fixed by the damping of rotating modes rather than by the slowest relaxation.
% CORE-END

\begin{acknowledgments}
OpenAI ChatGPT (GPT-5.6 family; final audit with GPT-5.6 Sol, accessed September 2026) was used under author-directed prompts to assist with literature synthesis, algebraic derivations, code review, numerical verification, and drafting. The author checked the derivations against the stated assumptions, reran the reported numerical tests from source code, reviewed the cited literature, and takes full responsibility for the content.
\end{acknowledgments}
\paragraph*{Data availability.}The analysis and plotting code, numerical checks, simulation design, and compact numerical outputs are publicly available in the versioned repository Ref.~\cite{CodeRepository}. Complete per-record counts and seeds for the prespecified power study are also included in the Supplemental Material. No experimental data were used.

\end{document}